\documentclass{webofc}
\usepackage[varg]{txfonts}   
\usepackage{graphicx}
\usepackage{subfigure}
\usepackage{placeins}
\usepackage{lmodern}
\usepackage{epsfig}
\usepackage{changepage}
\usepackage{wrapfig}
\usepackage{tikz}
\usepackage{tcolorbox}
\usepackage{xcolor}
\usepackage{epstopdf}
\usepackage{bm}
\newcommand\ignore[1]{}

\graphicspath{{C:/Users/Timothy/Desktop/Physics/Figs/}}

\usepackage{titlesec}
\titlespacing*{\subsubsection}{5pt}{.25\baselineskip}{.25\baselineskip}
\titlespacing*{\subsection}{5pt}{.25\baselineskip}{.25\baselineskip}
\titlespacing*{\section}{5pt}{.25\baselineskip}{.25\baselineskip}
\begin{document}
\title{Pomeron Physics at the LHC}
%
%

\author{\firstname{Federico} \lastname{Deganutti}\inst{1,2}\fnsep\thanks{\email{fedeganutti@ku.edu}} \and
        \firstname{David} \lastname{Gordo Gomez}\inst{1,3}\fnsep\thanks{\email{david.gordo@csic.e}} \and
		\firstname{Timothy} \lastname{Raben}\inst{1}\fnsep\thanks{\email{timothy.raben@ku.edu} {\bf Speaker}} \and
        \firstname{Christophe} \lastname{Royon}\inst{1}\fnsep\thanks{\email{christophe.royon@ku.edu}}
}

\institute{Department of Physics and Astronomy, University of Kansas, 1082 Malott, 1251 Wescoe Hall Dr., Lawrence, KS 66045-758 
\and
           Dipartimento di Fisica ed Astronomia, Universit{\` a} di Firenze, Piazza di San Marco, 4, 50121 Firenze FI, Italia 
\and
           Instituto de Física Teórica UAM/CSIC, Nicolás Cabrera 13-15, Campus de Cantoblanco UAM, 28049 Madrid, Espa{\~ n}a \& Universidad Aut{\' o}noma de Madrid, E-28049 Madrid, Espa{\~ n}a
          }

\abstract{%
We present current and ongoing research aimed at identifying Pomeron effects at the LHC in both the weak and strongly coupled regimes of QCD.
}
\maketitle
\section{Introduction}
\label{intro}
\FloatBarrier

Hadron collisions in many kinematic regimes display Regge behavior where cross sections grow as a power of the center of mass energy. Regge Theory grew out of pre-QCD S-matrix theory.  Amplitudes are seen as unitary, Lorentz invariant functions of analytic momenta.\footnote{The goal was to describe scattering experiments without a detailed description of an underlying microscopic theory.} Poles in scattering amplitudes represent particle exchange or bound state production.\footnote{Generically amplitudes also have cuts corresponding to multiparticle production and non-linear interactions.  The physics of cuts much less well understood and we omit any detailed discussion}  Using a partial wave analysis, the dominant contribution to simple amplitudes is the exchange of an \emph{entire trajectory} of particles: Pomeron exchange.  Single Pomeron exchange is predictive of a power-law rise in the total cross section\footnote{Asymptotically, as $s\rightarrow \infty$, cross-sections are likely subject to the Froissart bound indicating a maximal growth $\sigma_{tot}\sim log^2(s)$.} $\sigma_{tot} \sim s^{\alpha_0-1}$ Since the beginning of Pomeron physics, fits to p-p total cross section data have shown some power-law behavior.\footnote{The contribution responsible for the empirically observed $\sigma_{tot}$ rise is referred to as the \emph{soft Pomeron}.  Although recent $\sigma_{tot}$ measurements show a deviation from a strict power-law, the soft Pomeron approach is still used to fit data and for modeling.\cite{Menon:2013vka}} In these proceedings we review progress in using Pomeron exchange to describe LHC data.

\begin{figure*}[!ht]
\begin{center}
  \subfigure[]{
     \includegraphics[scale=.4]{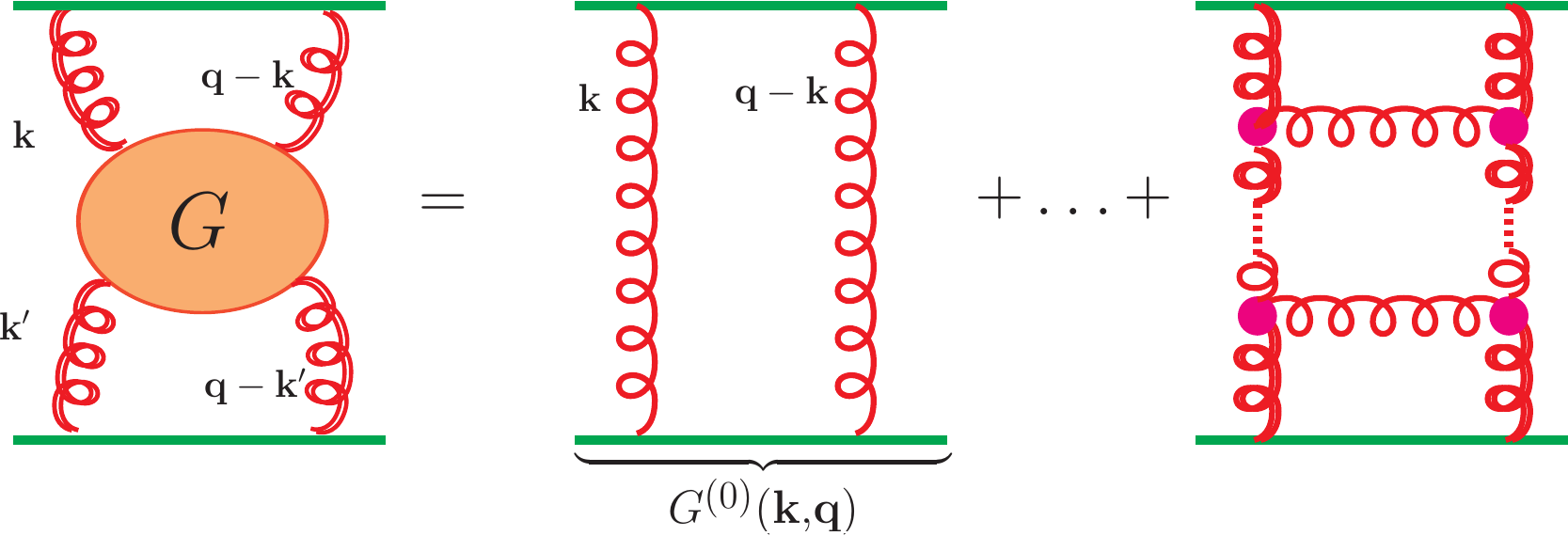}
  \label{fig:bfkl-ladder}
  }
  \hspace{35pt}
  \subfigure[]{
     \includegraphics[scale=.32]{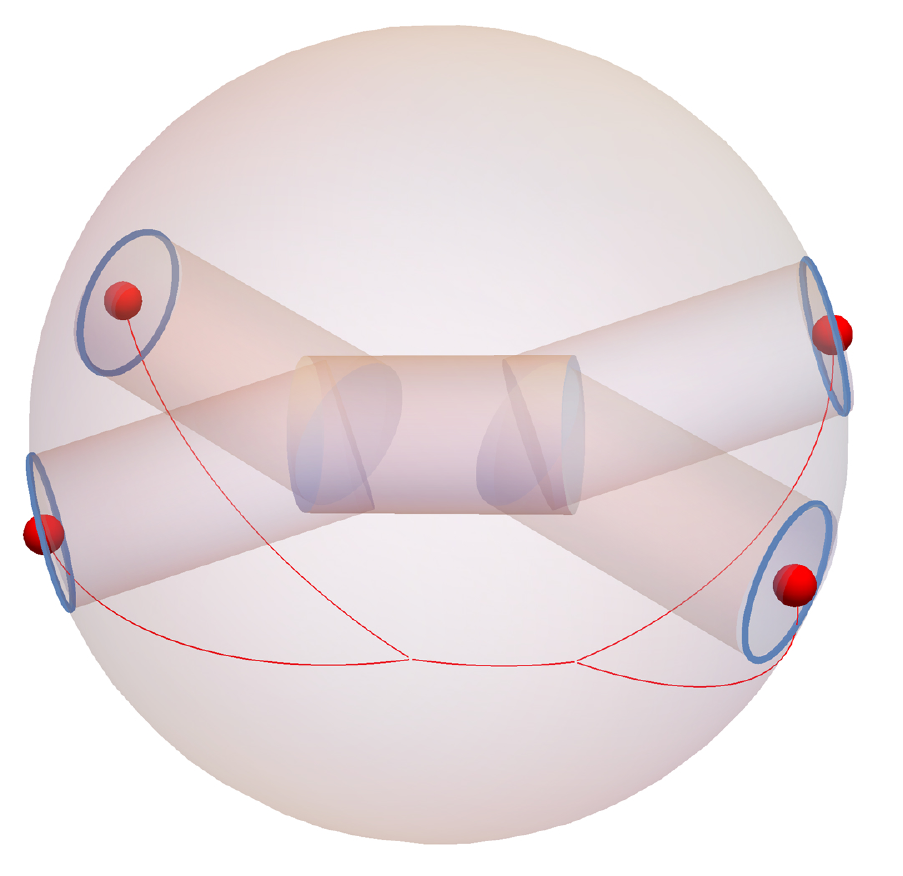}
  \label{fig:ads-cartoon}
  }
\end{center}
\vspace{-25pt}
\caption{\small (a) Ladder diagram for the BFKL weak coupling Pomeron. (b) Cartoon of a 2-to-2 scattering process using the AdS/CFT applicable for the BPST strong coupling Pomeron.\normalsize}
\label{fig:pomerons}
\vspace{-20pt}
\end{figure*}

\paragraph{Weak Coupling Pomeron} Balitsky, Fadin, Kuraev, Lipatov (BFKL) asked, and provided the first important answers to, the question: what happens in the Regge limit of QCD? Large logarithms get in the way of usual perturabtion theory; a resummation of terms $\alpha_s\,log(s)$ to all orders is necessary. Scattering amplitudes are dominated by the t-channel exchanged of an infinite ladder of reggized gluons (See Figure\ref{fig:bfkl-ladder}) leading to Regge behavior (power law growth of cross sections with energy). The BFKL equation, an integral equation for Green's function in Mellin space, is used to described this exchange kernel. The solution can be written
\small
\begin{equation}
G(\mathbf{k},\mathbf{k'},\mathbf{q},Y)=\int\limits_{-i\infty}^{+i\infty} \frac{d\omega}{2\pi i}e^{Y\omega}f_\omega(\mathbf{k},\mathbf{k'},\mathbf{q})\rightarrow \int\limits_{-i\infty}^{+i\infty} \frac{d\omega}{2\pi i}e^{Y\omega}
\sum\limits_{n \in \mathcal{Z}} \int\limits_{\frac{1}{2}-i\infty}^{\frac{1}{2}+i\infty} \frac{d\gamma}{2\pi\, i} \frac{E_{\gamma,n}(k)E_{\gamma,n}^*(k')}{\omega - \bar{\alpha}_s \chi (\gamma,n)} \, ,
\end{equation}
\normalsize
where in the leading log (LL) approximation: $ \chi(\gamma,n)=2\psi(1)-\psi(\gamma+\frac{|n|}{2})-\psi(1-\gamma+\frac{|n|}{2})$ and $\omega_0=\frac{4\alpha_s N_c}{\pi}ln(2)$.

\paragraph{Strong Coupling Pomeron}

Pomeron phenomenology is applicable over a wide range of kinematics and, particularly if one focuses on small momentum transfers, involves non-perturbative physics. The most promising description of a unified soft/hard Pomeron framework comes from the AdS/CFT duality conjecture. In this framework, a $\mathcal{N}=4$ susy Yang-Mills theory is dual to a string theory in a higher dimensional curved space.  The utility of such a duality is far beyond the scope here, but we mention a few pertinent properties: the duality is strong/weak so that a perturbative, weakly coupled string theroy calculation can give insights into strongly coupled Yang-Mills; the gluon sector of the $\mathcal{N}=4$ theory is similar to that of QCD; and the extra dimensions and new fields become natural mechanisms for modeling the onset of confinement and saturation.

Within the AdS/CFT duality, the Pomeron has been identified with the Regge trajectory of the graviton\footnote{The BPST Pomeron naturally connects a hard and soft regime and has diffusive behavior similar to the BFKL Pomeron. An initial description describing the first calculation of $1/\sqrt{\lambda}$ effects can be found in \cite{Kotikov:2004er}.}. (Otherwise known as the BPST Pomeron\cite{Brower:2006ea}) Scattering amplitudes computed via Feynman diagram like approach (See Figure\ref{fig:ads-cartoon} for a pictorial.) with the leading Regge amplitudes written as a convolution of wavefunctions and Reggeon propagators over AdS space: $A \sim \psi_1(z) \psi_2(z)\ast \chi(z,z',s,t) \ast \psi_3(z') \psi_4(z')$. Here s and t are Mandelstam like invariants and z is an AdS coordinate. Reggeon propagators are reminiscent of the weak coupling partonic description

\vspace{-5pt}\begin{equation}
\,\,\,\,\,\,\,\,\,\,\chi_R\sim \int dj (\alpha' \hat{s})^j(1+\text{cos}(-i\pi j))G_j(t,z,z')
\end{equation}
Operator dimensions of AdS Reggeons have an anomalous part and admit a non-trivial convergent expansion in terms of spin, coupling, and twist.  These $\Delta-J$ curves can be calculated to high order using a mix of conformal, string, and integrability techniques.  Minimizing these curves gives Reggeon intercepts.\cite{Brower:2014wha}

Applications of the BPST Pomeron have included deep inelastic total cross section measurements from HERA\cite{Brower:2010wf,Brower:2015hja,Brower:2014sxa}, vector meson production\cite{Costa:2013uia} and diffractive Higgs production \cite{Brower:2012mk} at the LHC, to predict glueball masses \cite{Ballon-Bayona:2015wra}, and to predict cross sections of non-diffractive central $\eta$ production \cite{Anderson:2016zon}.

\FloatBarrier

\section{Dijet Processes}\label{sec:bfkl}
\begin{figure*}[!ht]
\begin{center}
  \subfigure[]{
     \includegraphics[scale=.44]{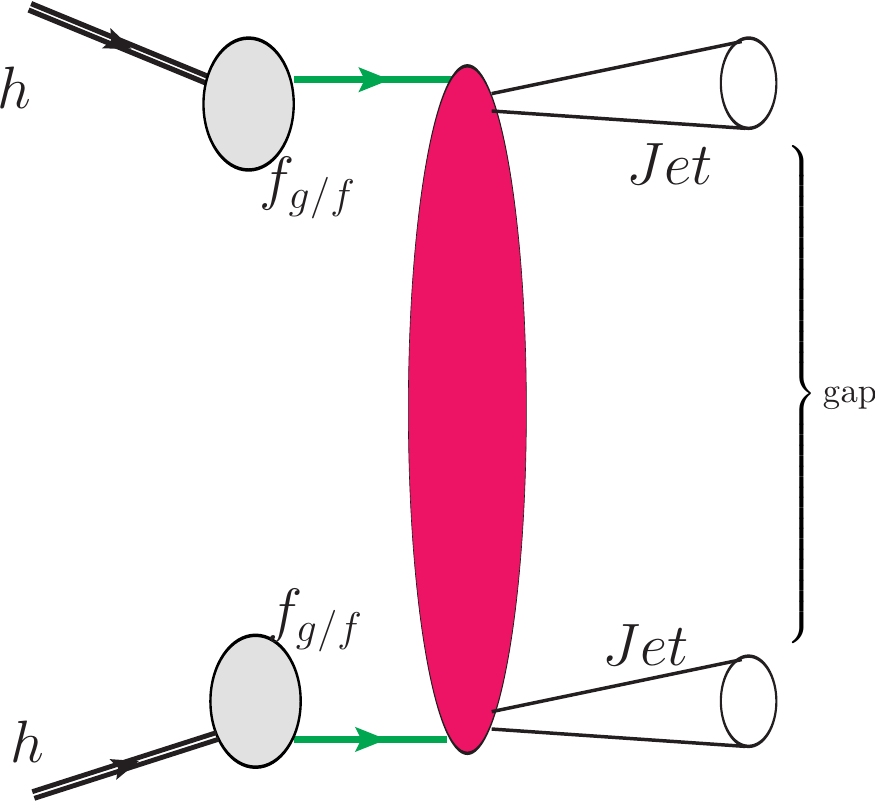}
  \label{fig:jgj}
  }
  \hspace{50pt}
  \subfigure[]{
     \includegraphics[scale=.4]{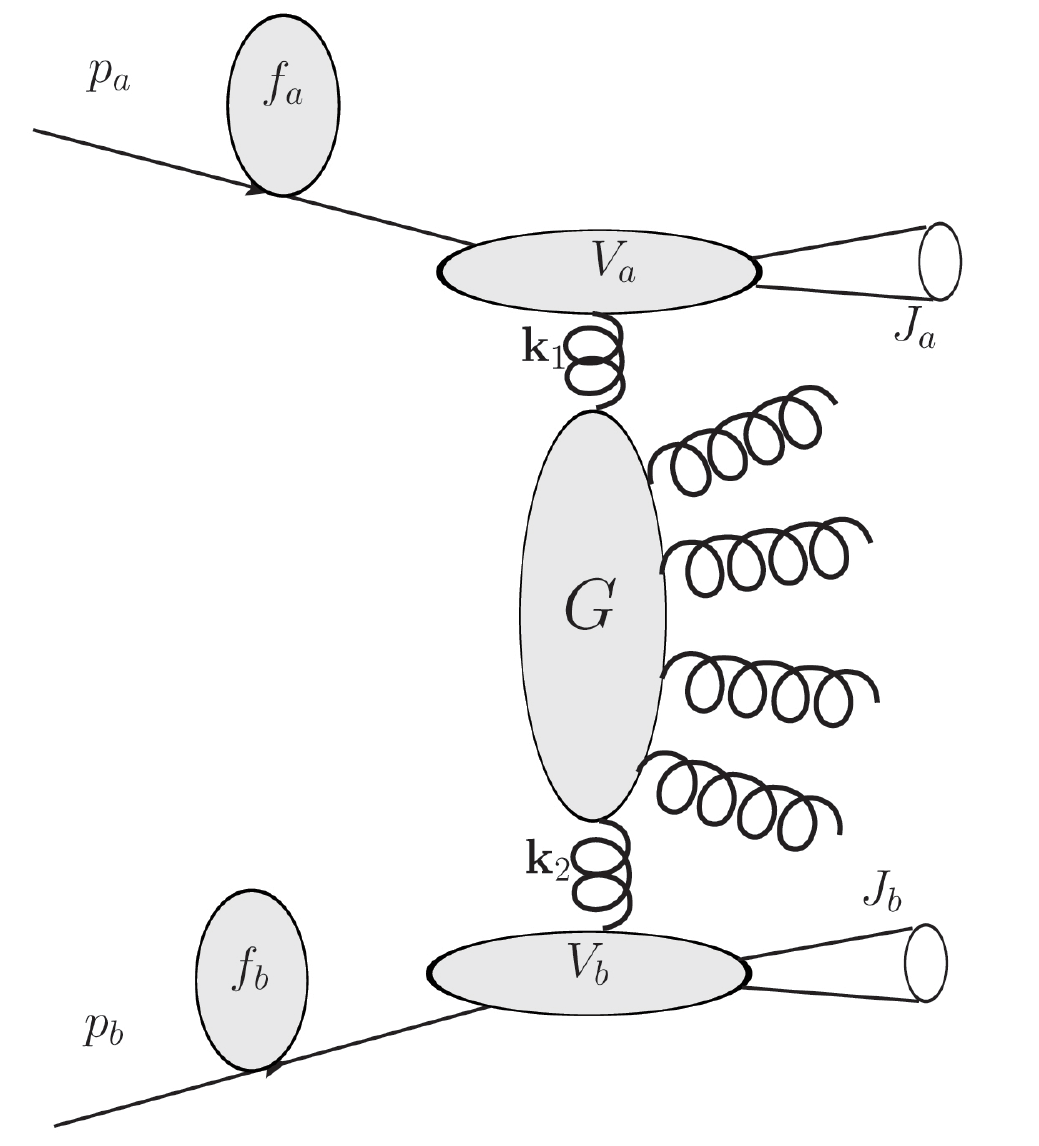}
  \label{fig:mn}
  }
\end{center}
\vspace{-25pt}
\caption{\small (a) Mueller-Tang jet-gap-jet process. (b) Mueller-Navelet inclusive dijet process. \normalsize}
\label{fig:dijets}
\vspace{-20pt}
\end{figure*}

Here we review previous efforts, and describe our program in progress, for observing BFKL effects at the current LHC collider kinematics\footnote{This kinematical regime is traditionally dominated by DGLAP evolution.}. We are focused on exclusive and inclusive dijet production as in Figure \ref{fig:dijets}.

\subsection{Mueller-Navelet Jets}
In the Mueller-Navelet (M-N)\cite{Mueller:1986ey} jet process all the radiation is treated inclusively except for the two jets farthest apart in rapidity. The link between this semi-inclusive process and the purely elastic amplitude of the Pomeron exchange is provided by the optical theorem.  Theoretically, this is one of the simplest jet processes sensitive to BFKL effects as the outer most dijet  can be thought of coming from 2-to-2 forward ($t=0$) parton scattering.  The BFKL equation greatly simplifies in this regime making both analytic and numerical computations much more tractable; the BFKL gluon ladder is seen as an interference diagram with real emissions that contribute to the cross section instead of being part of the  virtual corrections to the 2 to 2 scattering amplitude.

\ignore{
\begin{wrapfigure}{l}{0.33\textwidth}
\vspace{-15pt}
\begin{center}
\centerline{\psfig{figure=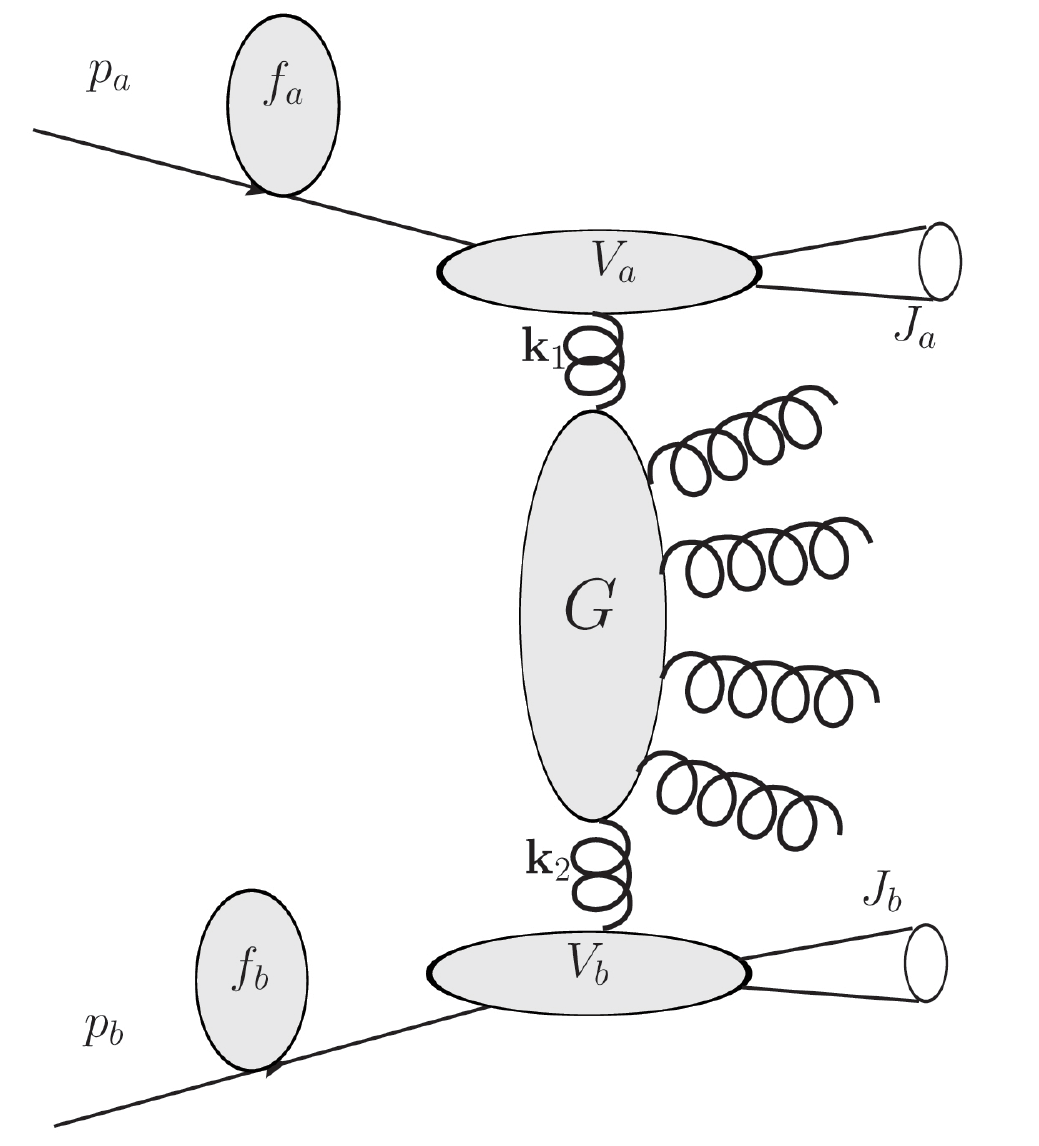,height=2.8in}}
\caption{\small Mueller-Navelet jet process. Tagged jets widely separated in rapidity with additional radiation in between. $G$ denotes the Green function and $V$ the jet vertex. \normalsize} 
\label{fig:F11}
\end{center}
\vspace{-30pt}
\end{wrapfigure}
}

BFKL dynamics are predicted to manifest as an increase in the  decorrelation of the momentum of the tagged jets. The configuration of perfect correlation, valid at tree level, is incrisingly altered by the addition of any other  emission.
In the forward Regge limit, the additional real emissions are enhanced by the large $\log{s}$. The larger is the rapidity separation between the outer most tagged jets, the wider is the rapidity interval that can be spanned by the ordered gluon emission, which generates the powers of logarithms: M-N jets with a large rapidity separation should be very sensitive to BFKL effects. The jet azimuthal (de)correlation can be quantified considering the Fourier modes of the average cosine of the angular azimuthal difference between the tagged jets $C_n/C_m=\langle\cos(n(\phi_{J_1}-\phi_{J_2}-\pi))\rangle/\langle\cos(m(\phi_{J_1}-\phi_{J_2}-\pi))\rangle$. Thanks to the high energy factorization, the coefficients $C_n$ are given by a convolution between the probe dependent jet vertices and the universal gluon Green function:

\begin{equation}
\begin{split}
\frac{C_{m}(| \mathbf{k}_{J_1}|,|\mathbf{k}_{J_2}|,Y)}{dY}\!=\!\int\! dx_1dx_2&f(x_1)f(x_2)dy_{J_1}dy_{J_2}\delta(y_{J_1}-y_{J_2}-Y)d\phi_{J_1}d\phi_{J_2}\\
\times\cos(m(\phi_{J_1}\!-\!\phi_{J_2}\!-&\!\pi))\int {d}\mathbf{k}_{1}\mathbf{k}_{2}V(\mathbf{k}_{J_1},x_{J_1},\mathbf{k}_{1})G(\mathbf{k}_{1},\mathbf{k}_{2},\hat{s})V(\mathbf{k}_{J_2},x_{J_2},\mathbf{k}_{2}).
\end{split}
\label{eq:MN}
\end{equation}

Here $Y$ is the rapidity difference between the outgoing jets.\footnote{Sometimes in the literature $\Delta y$ is used.} The remaining finite part is called the jet vertex $V(\mathbf{k}_{J_1},x_{J_1},\mathbf{k}_{1})$. The jet vertices, as for the gluon Green function, depend on the approximation order and can be defined unambiguously  in the BFKL approach.  They have been  proven to be infrared safe up to the next-to-leading order.

A test of the M-N BFKL predictions was carried out in 2013 by the CMS collaboration\cite{Khachatryan2016}\footnote{Similar types of analyses have been done with dijet inclusive processes at electron-ion colliders. For a recent example see\cite{Andreev:2015cwa}.}. Events with at least two emerging jets with with transverse energy $|\mathbf{k}_J| > 35$ GeV and $|y_J| < 4.7$ entered the analysis. Jets were reconstructed from the energy deposition in the calorimeter and clustered with the anti-$k_T$ algorithm\cite{Cacciari2008,Catani:1993hr} with a distance parameter $R=0.5$. Results were compared to the BFKL analytical predictions at NLL order\footnote{The NLO vertex found in the literature, and extensively used for theoretical analyses, is not in fully compatible with the M-N prescription that is used in experimental analyses. The inconsistency can be corrected, and the difference can amount up to $\sim 4-10\%$ change. Although important, it does not qualitatively alter the conclusions drawn in the analyses of CMS data. 
The corresponding paper is still unpublished\cite{fede}, but an outline can be found in\cite{Colferai:2017bog}.} and the predictions of several Monte Carlo generators like PYTHIA\cite{Sjoestrand2007} and HERWIG++\cite{Bahr:2008pv} which are based on the DGLAP evolution.

\begin{figure}[htb]
  \centering
	\includegraphics[width=.75\columnwidth]{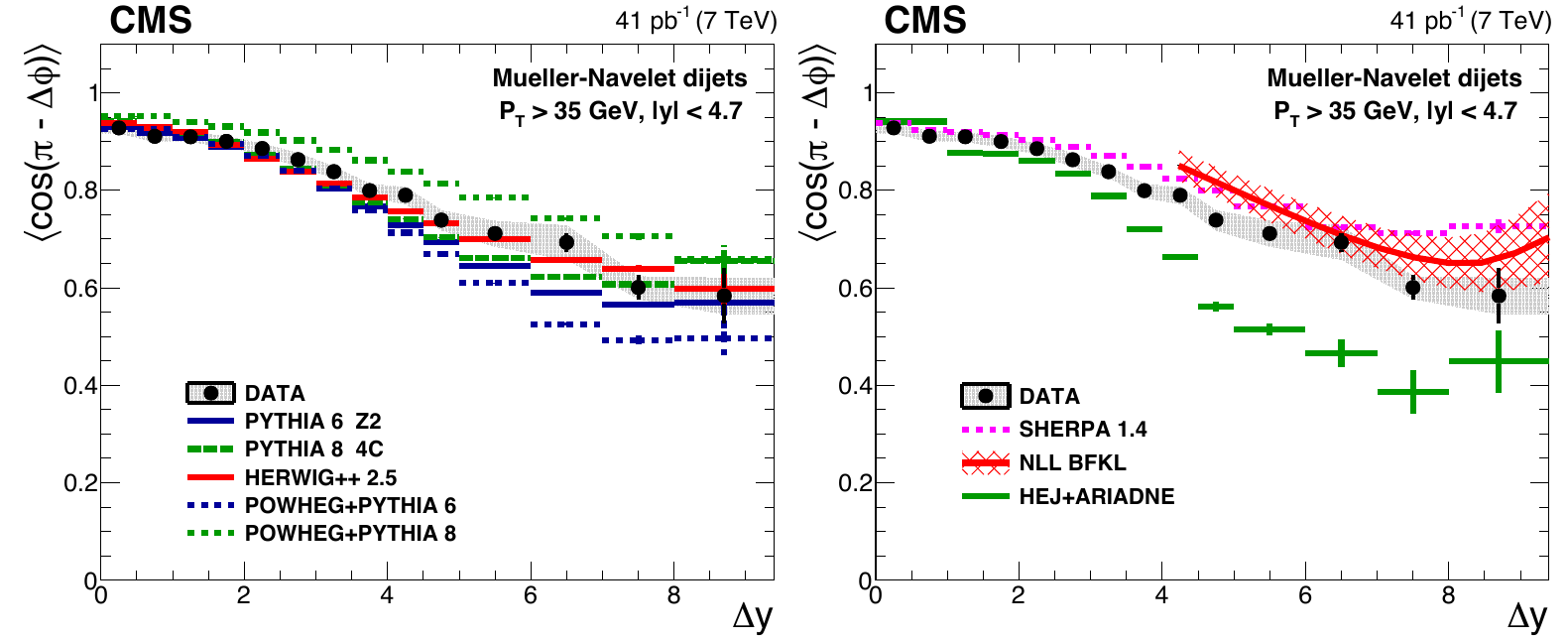}
	\caption{\small From \cite{Khachatryan2016}, average $\cos(\Delta \phi$), i.e. $C_1/C_0$, as a function of the rapidity difference $Y$ compared to DGLAP MC generators on the left and to the analytic NLL BFKL calculations at partonic level on the right. Similar analysis holds for most other ratios $C_m/C_n$. \normalsize}
     \label{FSQ12}
     \vspace{-20pt}
\end{figure}

Concerning the BFKL predictions, few conclusions were drawn: the analytic NLL calculation predicts a decorrelation above the observed data for the  three conformal moments in the whole range of rapidities, except for the data points at $Y\simeq 9$. A better agreement with data was observed for the ratios $C_n/C_m$ ($m,n\neq 0$) along the whole rapidity range. The DGLAP-based MC generator HERWIG++ reproduces the experimental data with satisfactory accuracy for all the observables.

\ignore{
\begin{figure}[htb]
	\centering
\includegraphics[width=1.0\columnwidth]{FSQ_3.pdf}
    \caption{\small Coefficients ratios as a function of $Y$ compared to DGLAP inspired MC generators on the left and to the analytic NLL BFKL  calculations at partonic level on the right. Similar analysis holds for the $C_3/C_2$ ratio.\normalsize}
  \label{FSQ34}
\end{figure}
}

The analysis confirms that contamination from DGLAP evolution dijets affects substantially all the observables containing the total cross-section $C_0$ and suggests that a description of such observables based on the BFKL approach alone is perhaps not feasible at the current initial energies; the kinematic domain explored in the CMS analysis seems to lie in a transition region between the DGLAP and BFKL regimes for angular decorrelation. This opens the path for alternative or complementary strategies to better disentangle BFKL effects on top of the standard M-N angular decorrelation observables.  

\subsection{New Observables for Inclusive Dijet Production}\label{sec:mnjets}

\begin{figure*}[!ht]
\begin{center}
  \subfigure[]{
     \includegraphics[scale=.52]{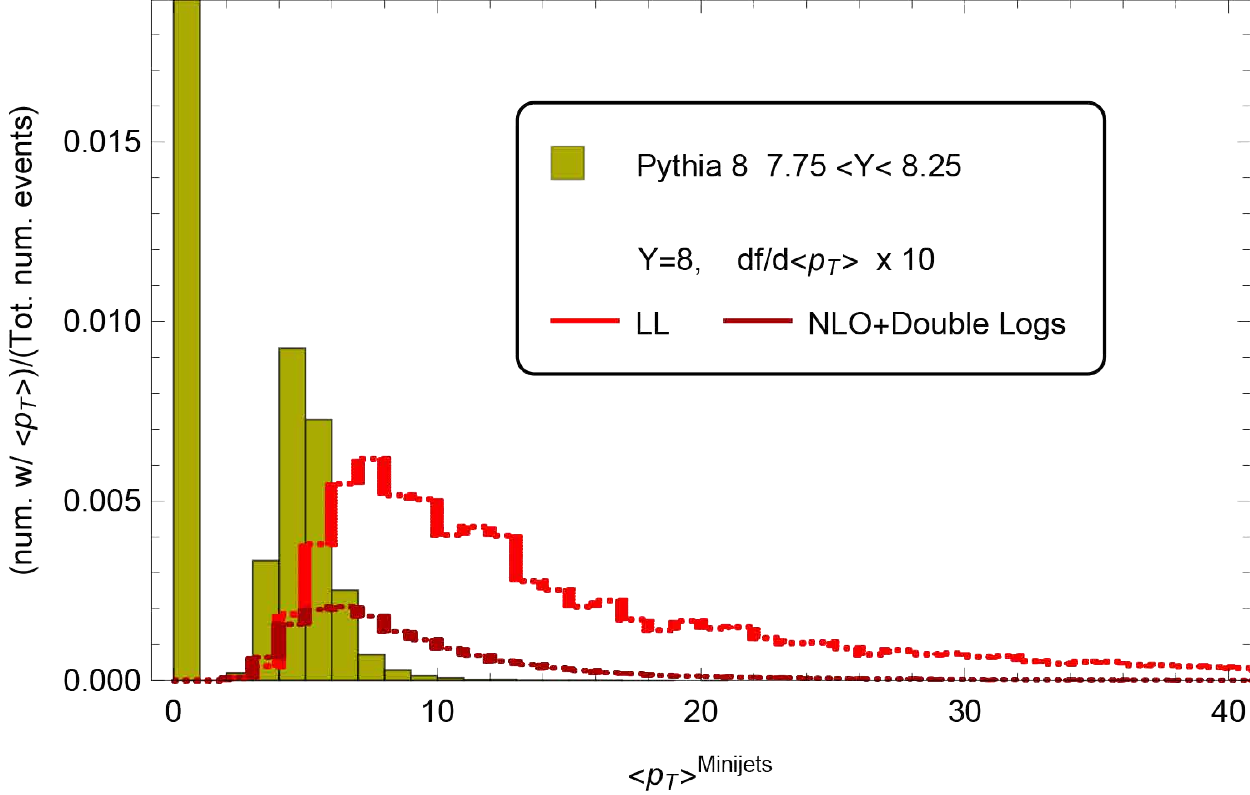}
  \label{fig:pt8}
  }
  \hspace{5pt}
  \subfigure[]{
     \includegraphics[scale=.52]{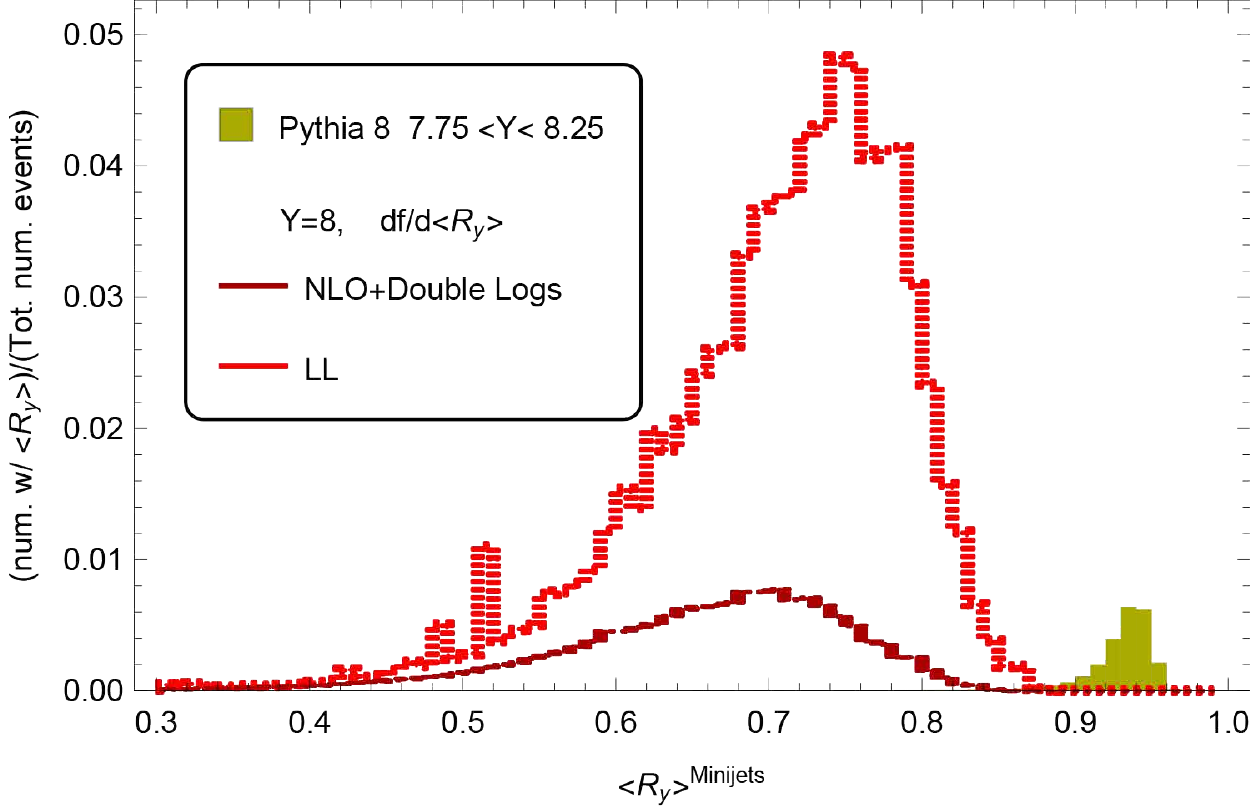}
  \label{fig:ry8}
  }
\end{center}
\vspace{-25pt}
\caption{\small Preliminary comparison of mini-jet $p_\mathrm{T}$ and $\mathcal{R}_y$ between BFKLex (red curves) and DGLAP-based PYTHIA8 (gold). PYTHIA results have been rescaled by one order of magnitude for ease of comparison. BFKLex data is from \cite{Chachamis:2016okn}.\normalsize}
\label{fig:pt}
\vspace{-20pt}
\end{figure*}

Since the geometry of the CMS detector limits the highest accessible rapidity, BFKL effects cannot be enhanced at LHC via increasing the center-of-mass energies. However, observables could be more sensitive to the differences between DGLAP and BFKL evolutions. A promising candidate is to consider the same Mueller-Navelet definition with an additional requirement that excludes events with an insufficient number of distinguished jets in between the tagged jets: counting mini-jet radiation. In fact, the BFKL ordered emission is expected to give rise to a larger mini-jet multiplicity compared to the collinear emissions re-summed in the DGLAP picture. 

In addition to multiplicity, other observables were recently proposed \cite{Chachamis:2015ico,Chachamis:2016rxo} using this mini-jet radiation
\begin{equation}
\langle p_T\rangle = \frac{1}{N} \sum_i |k_i|,  \qquad \langle \theta\rangle = \frac{1}{N} \sum_i \theta_i, \qquad \langle \mathcal{R}_y\rangle = \frac{1}{N+1} \sum \frac{y_i}{y_{i-1}},
\end{equation}
reflecting the average transverse momentum, emission angle, and rapidity ratio of the mini-jets respectively.  Initial investigations showed that these observables could be simulated using the BFKLex MC to give non trivial predictions. In Figure \ref{fig:pt}, results of a comparison between DGLAP based PYTHIA8 and the BFKLex for the average transverse momenta and rapidity ratio are plotted.  These preliminary results\cite{bfklvsdglap} are extremely encouraging: the BFKL average $p_T$ has a very heavy tail and the BFKL $\mathcal{R}_y$ is peaked differently from the DGLAP expectation.

However, there are still challenges left to pursue.  The data in Figure \ref{fig:pt} has cuts for the M-N outer jets similar to that of CMS, $|p_\mathrm{T}|>35$ GeV.  However, the mini-jets are simulated with a rather modest $|k_i|>1$ GeV; increasing to the sensitivity of CMS will require much longer runs.  Secondly the overall normalization of the BFKLex is \emph{not} fixed.  This is primarily an effect of not incorporating PDF effects in the MC; BFKLex only uses the BFKL kernel.  This issue can be circumvented either by an including PDF effects in BFKLex or normalizing the distributions in a region where there is a small overall rapidity gap and small $p_T$ difference between the outer most jets; a region where both DGLAP and BFKL effects might be suppressed and the predictions should agree.  The consideration of new observables, like higher moment companions of $\mathcal{R}_y$ might also alleviate these issues.

\subsection{Mueller-Tang Jets}\label{sec:jgj} 
 
\begin{figure}
    \hspace{40pt}\includegraphics[width=0.23\textwidth]{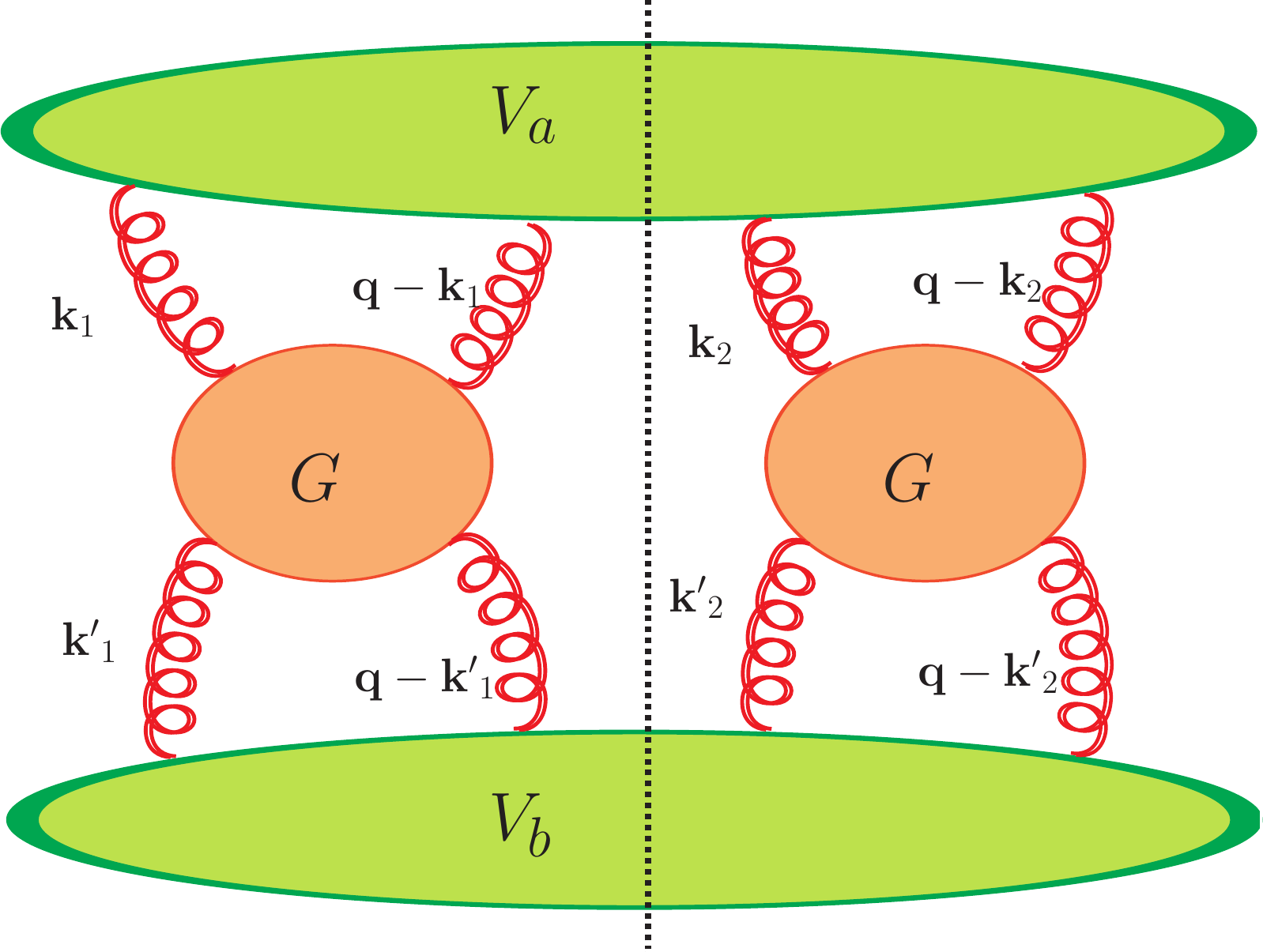}\\
\centering
    \sidecaption
   \includegraphics[width=0.6\textwidth]{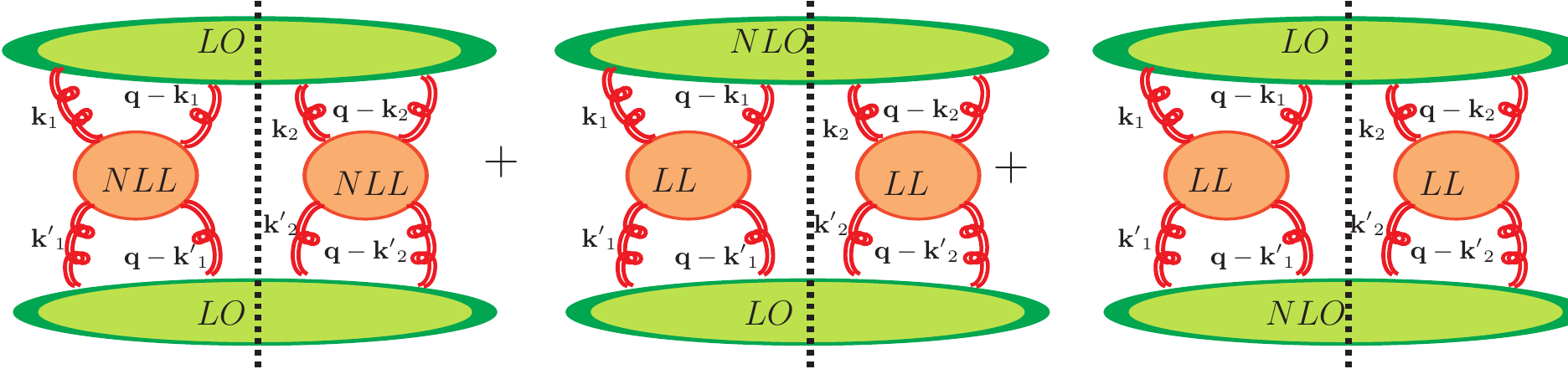}
\caption{(Top) Schematic representation of the partonic elastic scattering via color-singlet exchange. $G$ denotes the Green function and $V$ the impact factors. (Bottom) Combinations of jet vertex and Green's function order that contribute to the full NLO M-T process.}
\label{fig:nlo}
\vspace{-20pt}
\end{figure} 

\begin{figure*}[!ht]
\vspace{-10pt}
\begin{center}
 \subfigure[]{
      \includegraphics[scale=.8]{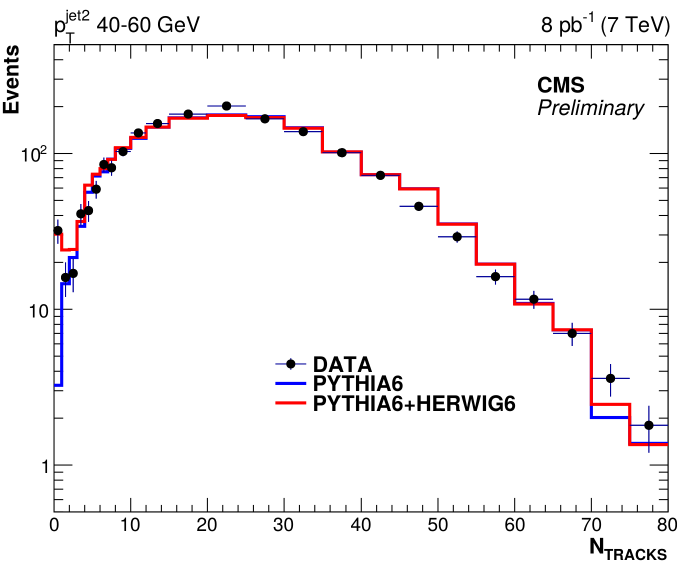}
  \label{fig:TRACKS}
  }
  \hspace{30pt}
\subfigure[]{
     \includegraphics[scale=.55]{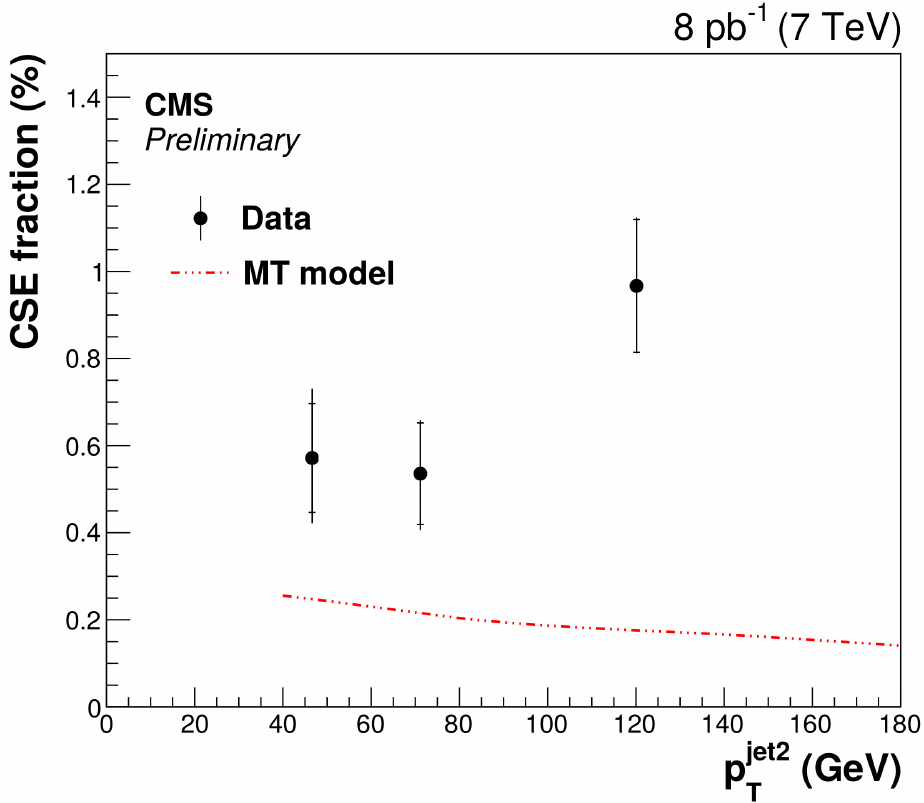}
  \label{fig:CSE}
  }
\end{center}
\vspace{-25pt}
\caption{ \small From \cite{cms}: (a) Number of charged particle tracks $N_\mathrm{Tracks}$ for $40$ GeV$<p_\mathrm{T}^\mathrm{sublead}<60$ GeV with $|\eta^{1,2}|>1.5$ and $\eta^1\cdot\eta^2<0$ in the fixed rapidity interval $|\eta|<1$ of CMS. The red curve combines the LL  M-T dijet and PYTHIA6 NLO dijet events. The excess at low multiplicities can be explained due to the color-singlet exchange. (b) The fraction of color-singlet exchange $f_\mathrm{CSE}$ dijet events as a function of the subleading $p_\mathrm{T}$ measured for $\sqrt{s}$=7 TeV, compared to the LL BFKL predictions for $N_\mathrm{Tracks}=0$. A large numeric and qualitative disagreement between the two can be seen. \normalsize}
\label{fig:FSQ}
\vspace{-20pt}
\end{figure*}

The process of using large rapidity gaps to isolate and measure high energy asymptotic behavior goes back to ~\cite{Bjorken1993a,Dokshitzer:1987nc,DOKSHITZER1992116}. The properties of the BFKL hard Pomeron at finite momentum transfer can be investigated at hadron colliders by looking for highly exclusive processes where a dijet is separated by a large rapidity interval devoid of radiation ~\cite{Mueller:1992pe}. These dijet events are known as Mueller-Tang dijet events or jet-gap-jet events \cite{Mueller1992}. At the parton level the simplest configuration consists of 2-to-2 elastic parton scattering where, at large s, the non-forward elastic amplitude is dominated by the exchange of a Pomeron. No real emission is allowed in the internal rapidity region suppressing DGLAP evolution\footnote{This can be seen in Figures\ref{fig:TRACKS} and \ref{fig:mult} in the large excess of BFKL events at zero multiplicity.}. Although this is a much cleaner experimental signal of BFKL physics, handling the non-forward amplitude is a more difficult task theoretically.

An observable that gained large popularity is the differential cross section ratio of jet-gap-jet events to inclusive dijet events: $R = d\sigma^{JGJ}/d\sigma^{dijet}$. In the BFKL approach the jet-gap-jet partonic cross section can be computed as\footnote{The observable $R$ contains the gap survival probability $\mathcal{S}$.  This has an intrinsic non-perturbative nature \cite{Levin1998,Levin1999,Levin2012,Ryskin2008,Chuinard:2015sva}  In most applications it has been assumed to be an empirically determined constant depending only on the center of mass energy, although recent results have shown it should have some kinematic dependence\cite{Luszczak:2016csq,Babiarz2017}.} 
\begin{equation}
\begin{split}
\frac{d\hat{\sigma}}{dJ_1dJ_2d\mathbf{q}}=\int d\mathbf{k}_{1,2}d&\mathbf{k}'_{1,2}V_a(\mathbf{k}_1,\mathbf{k}_2,J_1,\mathbf{q})\mathcal{G}(\mathbf{k}_1,\mathbf{k}'_1,\mathbf{q},Y)\times\\
&\mathcal{G}(\mathbf{k}_2,\mathbf{k}'_2,\mathbf{q},Y)V_b(\mathbf{k}'_1,\mathbf{k}'_2,J_2,\mathbf{q}),
\end{split}\label{eq:MT}
\end{equation}
where $J=\{\mathbf{k}_J,x_J\}$ collects the variables that specify the jets.
It is represented schematically in Fig. \ref{fig:nlo}. The non-forward gluon Green's function and the jet vertices depend on the approximation order\footnote{If the jet vertices are kept at leading order the general expression can be \emph{greatly} simplified. This fact has been extensively used in the phenomenological analyses.}.

\begin{figure}
\centering
\sidecaption
\includegraphics[width=0.45\textwidth]{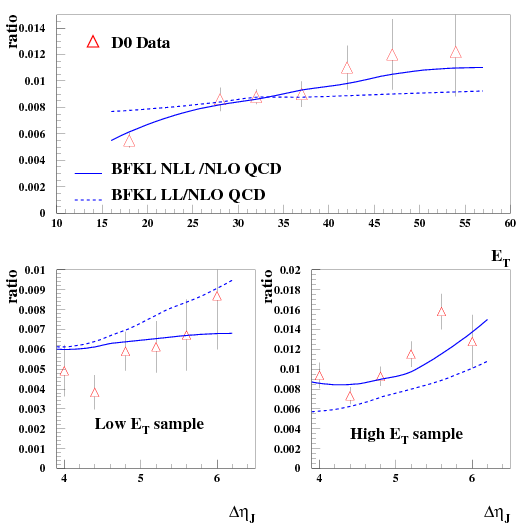}
	\caption{\small Comparisons between the D0 measurements of the jet-gap-jet event ratio using the NLL-BFKL kernel (solid line), LL-BFKL (dashed line) predictions with respect to the 2-to-2 NLO DGLAP prediction. Taken from \cite{Kepka:2010hu}. The NLL calculation is in fair agreement with the data while the LL one leads to a worse description. \normalsize}
	\label{fig:D0}
	\vspace{-20pt}
\end{figure}

To identify M-T events, the two hardest jets resulting from the color-singlet exchange are selected which: are strongly correlated in their azimuthal angular separation; are balanced in transverse momenta; and lie on opposite hemispheres of the central detector ($\eta_1 \cdot \eta_2 < 0$). Experimentally, the rapidity gap is defined by means of the absence of particle activity in a fixed rapidity region in the central detector, for instance, at the Tevatron  they counted the number of calorimeter towers with energy $E>$ 0.2 GeV in the fixed rapidity region $|\eta|<1$ and defined the rapidity gap events as the ones falling in the lowest multiplicity. It is possible to obtain a high-purity sample of jet-gap-jet events with this definition; for a judicious choice of jet momentum, rapidity, and size gap-the DGLAP evolution is highly suppressed.
\ignore{
The production cross section of two jets with a gap in rapidity between them reads~\cite{Marquet2012,Kepka2010a}
\begin{equation}
\frac{d \sigma^{pp\to XJJY}}{dx_1 dx_2 dp_\mathrm{T}^2} ={\cal S}f_{\mathrm{eff}}(x_1,p_\mathrm{T}^2)f_{\mathrm{eff}}(x_2,p_\mathrm{T}^2)
\frac{d \sigma^{gg\rightarrow gg}}{dp_T^2}
\label{jgj}
\end{equation}

where $p_\mathrm{T}$ is the transverse momentum of the two jets, $x_1$ and $x_2$ their longitudinal fraction of momentum with respect to the incident hadrons, $S$ is the rapidity gap survival probability --which in general has a dependence on the rapidity and transverse momenta of the jets--, and $f_\mathrm{eff}$ are the effective parton distribution functions~\cite{Kepka2010a}.

 }

M-T jets were investigated at the Tevatron at $\sqrt{s}=1.8$ TeV\cite{1998,PhysRevLett.80.1156} and by the CMS collaboration at 7 TeV\cite{cms}\footnote{This is the only experimental analysis coming from the LHC to date, though there is an ongoing 13 TeV analysis according to private communications with members of the CMS collaboration.}. The rapidity gap condition was applied only on the central rapidity region $|\eta|<1$ and only on charged particles with $p_\mathrm{T}>0.2$ GeV. The ratio of dijet events with rapidity gap to the total of dijet events has been measured as a function of the second leading jet $p_\mathrm{T}$ and the rapidity difference $\Delta\eta$. Even in this regime, the BFKL can be used to compute the partonic elastic amplitude. The implementation in a Monte Carlo is necessary in order to make a comparison with data; cross section measurements are sensitive to the jet size and the observed rapidity separation between the leading jets is smaller due to the soft radiation from the edge of the jets. Underlying event effects, which are estimated to be small, with the latest MC tuning should also be incorporated in the simulation. A phenomenological study was done of data released by the D0 Collaboration\cite{Kepka:2010hu} to the BFKL prediction by implementing the NLL BFKL kernel into HERWIG6.5. (See Figure\ref{fig:D0}). Importantly it was found that improving the BFKL kernel to NLL order and including a large number of conformal spin contributions are both necessary to obtain good agreement with data.\ignore{\footnote{HERWIG calculates the dijet events using a tuned DGLAP approach.}} In the CMS analysis summary, the results are compared only with the LL BFKL predictions (See Fig. \ref{fig:CSE}). These plots hint the need of higher-order corrections.

\ignore{
\begin{figure*}[ht]
\vspace{-25pt}
\begin{center}
 \subfigure[]{
     \includegraphics[scale=.4]{jgt-pom.pdf}
  \label{fig:singlepom}  
  }
  \hspace{45pt}
 \subfigure[]{
     \includegraphics[scale=.3]{Figure_002.jpg}
  \label{fig:dblpom}  
  }   
\end{center}
\vspace{-10pt}
\caption{\small (a) Schematic diagram of a M-T event in a $pp$ collision. The partons scatter elastically due to Pomeron exchange, the latter inducing the rapidity gap between the hardest two jets. (b) Topology of the jet-gap-jet events in the $\eta-\phi$ plane. \normalsize} 
\label{fig:jgj-diagrams}
\end{figure*}
}

The hard pomeron exchange embedded into HERWIG6.5 lacks the high order jet vertex correction to complete the NLL order. Nonetheless, it is expected that once the normalization is fixed for a given data set \footnote{The rescattering of proton remnants can destroy the gap resulting in a suppression factor that can affect at least the overall normalization.} the NLL BFKL is able to describe reasonably well the Tevatron data.

The charged multiplicity distribution measured by CMS in Fig. \ref{fig:TRACKS} is well described by DGLAP models (PYTHIA6) except at zero multiplicity, where we expect to see the jet-gap-jet events. The excess is consistent with the prediction of the HERWIG6.5 event generator using the color-singlet exchange process, even if the LL approximation BFKL kernel is used.
 There is a hint that the Mueller-Tang jets is an excellent venue to extract the signal of a BFKL Pomeron, even if the event generation was done with the LL approximation kernel and the leading conformal spin. Refining the theoretical prediction up to the full NLL BFKL order is an essential step toward this goal to show that the hard color-singlet exchange gives the correct differential distributions.

\subsection{Jet-Gap-Jet Predictions at NLO}\label{sec:mtjets}

\begin{figure*}[!ht]
\vspace{-10pt}
\begin{center}
 \subfigure[]{
     \includegraphics[scale=.28]{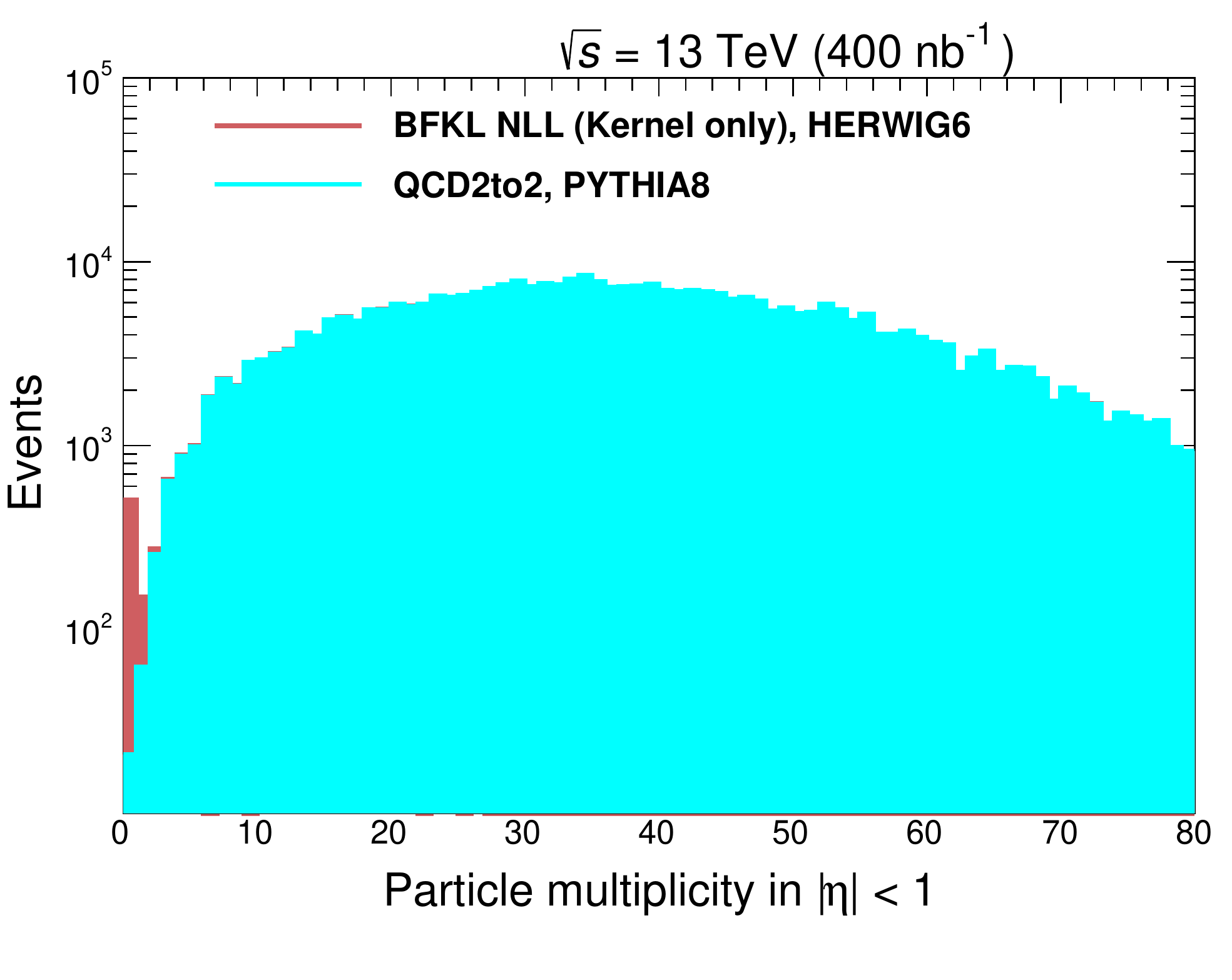}
  \label{fig:mult}
  }
  \hspace{30pt}
\subfigure[]{
     \includegraphics[scale=.28]{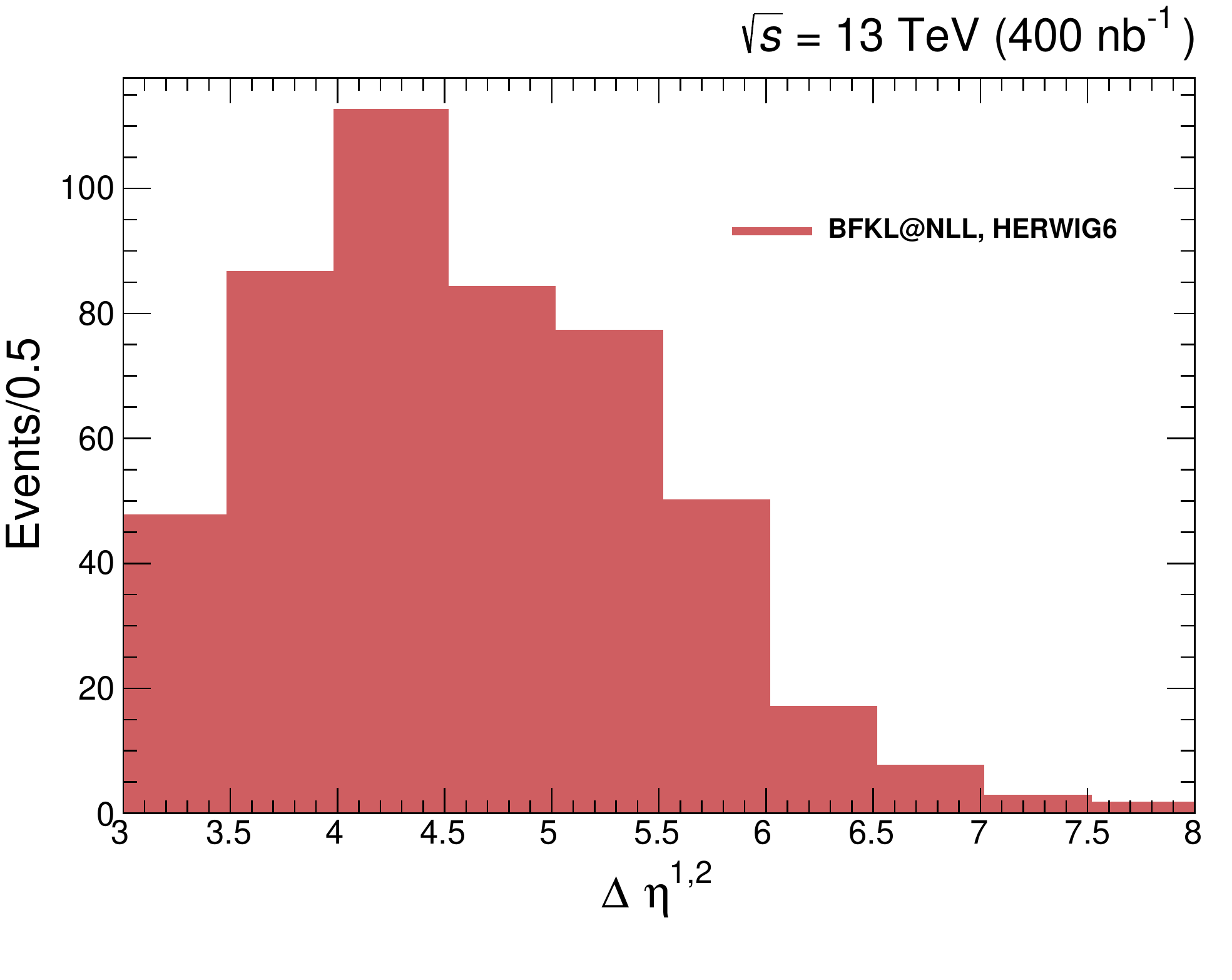}
  \label{fig:deta}
  }
\end{center}
\vspace{-25pt}
 \caption{ \small (a) Particle multiplicity in the fixed rapidity region $|\eta|<1$. Stable particles pass the selection cut $p_\mathrm{T}>$0.2 GeV. A high-purity Mueller-Tang dijets sample can be extracted in the lowest multiplicity bin. (b) Rapidity separation between the two hardest jets in the lowest multiplicity bin. \normalsize}
\label{fig:jgj-herw}
\vspace{-20pt}
\end{figure*}

Until now the description of M-T jets has focused on incorporating the NLL BFKL kernel, considering higher conformal spin corrections, and understanding soft events that can affect the rapidity gap signature.~\cite{Marquet:2012ra,Kepka:2010hu,Motyka:2001zh,Hentschinski:2014lma}. The path forward\footnote{A group of Uppsala University has studied the jet-gap-jet process since the time of the Tevatron \cite{Enberg2001,Ekstedt:2017xxy}. They have developed the HARDCOL package, a modified version of the PYTHIA6 event generator, where they embedded the solution of the non-forward LL BFKL equation for the jet vertices and kernel. This implementation considers also the multi parton-parton interactions with the latest tuning in PYTHIA6 based on LHC data at 7 TeV. In addition, they used the Soft Color Interaction model for color rearrangements in the final state through soft gluon exchanges, since such rearrangements can have large effects on rapidity gaps. This approach is complimentary to that proposed here.} is clear: add the NLO jet vertex\footnote{Recently the NLO jet vertices were calculated\cite{Hentschinski2014d,Hentschinski2014e} using Lipatov's effective action. Although the vertices were shown to be finite, there is a careful cancellation of soft and collinear divergences between the real and virtual corrections.  In addition, the jet definitions involve complicated non-analytic pieces to prevent gap contamination. Both of these require careful numerical treatments.} to the NLL BFKL kernel and perform a full NLO analysis of LHC data. Still, the full NLO calculation is numerically very non-trivial. The convolutions over the transverse momenta in Eqn. \ref{eq:MT} can be computed analytically if the vertices are  kept at the LL approximation. At NLO, they have to be solved numerically utilizing Monte Carlo integration techniques. Similarly to the M-N process, the corrections are expected to be large and important to accurately predict BFKL physics, though the calculation is absolutely necessary to verify this. In ~\cite{Kepka:2010hu}, to speed-up  computation time, the amplitude is fit to an ad-hoc parameterization and fed into HERWIG6.5 in place of standard QCD 2-to-2 process. More recently a variety of parameterizations was tested\cite{Trzebinski2015} and preliminary results show that the best fit parameterization can be physically motivated by considering asymptotics of the NLL kernel.  Figure \ref{fig:jgj-herw} was generated using this parameterization. Jets are reconstructed using the anti-$kt$ algorithm with a distance parameter of $R=0.4$ and the selection cuts described in the CDF, D0 and CMS analyses are applied to them. A definition of the rapidity gap similar to that used at the Tevatron and the LHC is used. While HERWIG6.5 counts with underlying event effects, the modeling does not take into account LHC-era results.

\subsection{Inclusive Central Production}
\label{sec:inc}
\begin{figure}
\centering
\sidecaption
    \includegraphics[width=0.4\textwidth]{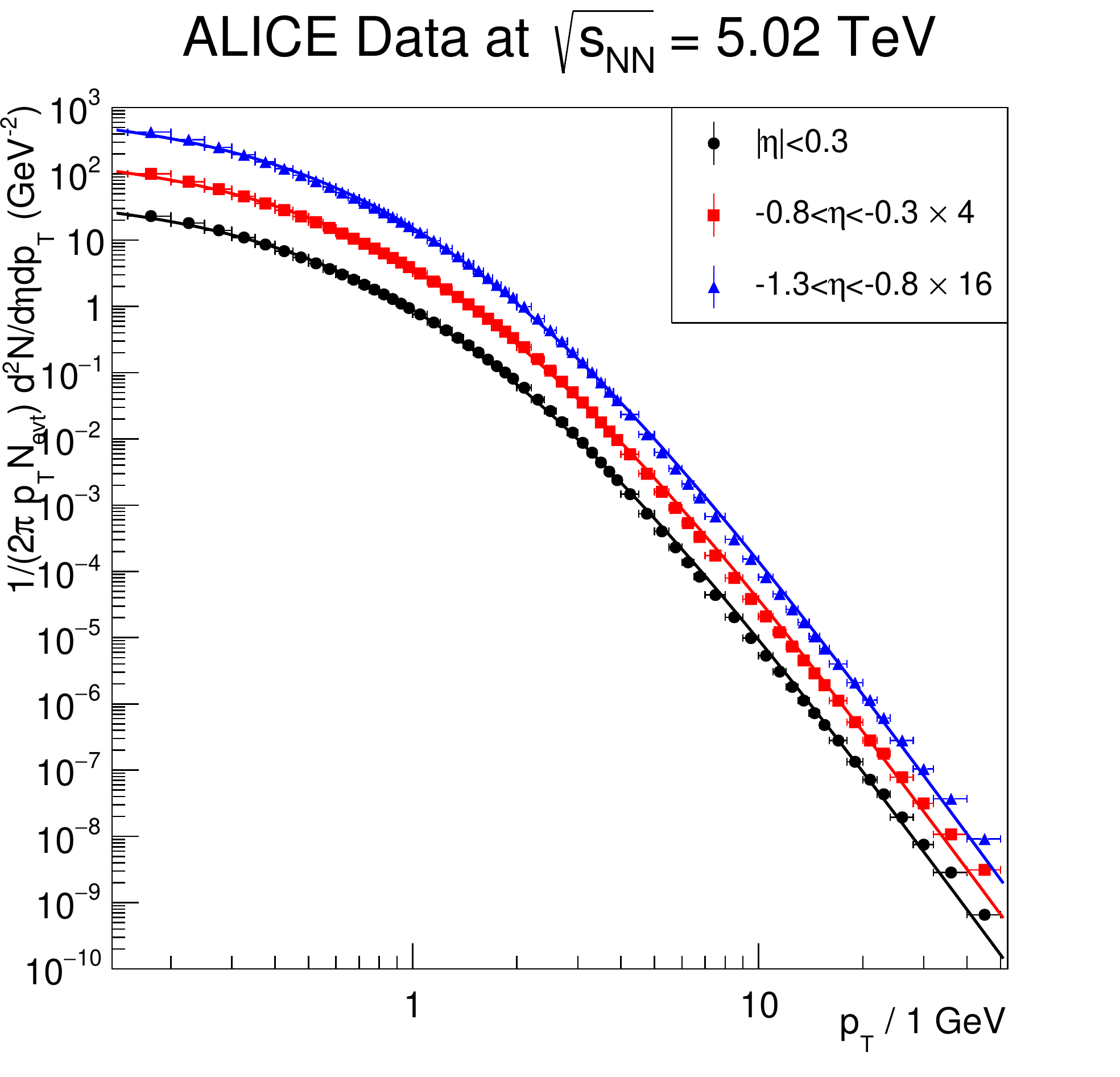}
  \caption{Fits of the ansatz, Eq.(\ref{eq:ansatz}), to the ALICE $\sqrt{s_{NN}}=5.02$TeV data set for the inclusive production of charged particles in the central regions.}
\vspace{-20pt}
\end{figure}  
The BFKL Pomeron, BPST Pomeron, and QCD as extremely short distances exhibit conformal symmetry\footnote{It is important to note that they are different symmetries.}. However, conformal symmetry has historically eschewed a description of scattering process because of the inability to define asymptotic states. Using the AdS/CFT, it was recently shown\cite{Nally:2017nsp,Nally:2017uep} that effects due to  conformal properties can be exhibited in LHC scattering.  The central idea, extended in these works, is that instead of considering the scattering of asymptotic states, the emphasis should be on the flow of IR safe quantities.  In addition, by utilizing generalized optical theorems, cross sections can be expressed as a discontinuity over the appropriate amplitude. For example, the differential cross section for inclusive central production can be computed via the AdS/CFT 
\begin{equation}\label{eq:ansatz}
\frac{1}{\sigma_{tot}}\frac{d^3\sigma_{ab\rightarrow X}}{d\bm{p}^2_{\perp}dy}\sim \frac{1}{2i\,s}Disc_{M^2>0}[\mathcal{T}_{abc'\rightarrow a'b'c}]\sim \frac{A}{(\bm{p}_{\perp}+C)^B}
\end{equation}
Using this ansatz, data from $\sqrt{s_{NN}}=5.02$ TeV p-pb collisions at ALICE, and 8 and 13 TeV p-p collisions at ATLAS were analyzed (See Figure \ref{tab:inc}). Results show conformal behavior near $B=2\Delta=8$ regardless of confinement model used. A lower cutoff $C\sim \Lambda_{qcd}$ is consistent with the onset of confinement effects. Further discussion of deviations from exact conformal behavior can be found in \cite{Nally:2017nsp}.

\begin{table}
\centering
\caption{Fitting parameters for holographic description of inclusive central production in p-X scattering.}

\resizebox{.75\columnwidth}{!}{
\begin{tabular}{cccc}\hline \label{tab:inc}
Dataset & A/10 (GeV$^{-2}$) & B  & C/(1 GeV)  \\\hline
ALICE 5.02 TeV, $\left|\eta\right|<0.3$  & 38.48 $\pm$ 8.26 & 7.23 $\pm$ 0.09 & 1.32 $\pm$ 0.04\\\hline
ALICE 5.02 TeV, $-0.8 < \eta < -0.3$   & 37.60 $\pm$ 7.97 & 7.22 $\pm$ 0.08 & 1.30 $\pm$ 0.04  \\\hline
ALICE 5.02 TeV, $-1.3 < \eta < -0.8$   & 43.00 $\pm$ 9.29 & 7.30 $\pm$ 0.09 & 1.31 $\pm$ 0.04  \\\hline
ATLAS 8 TeV & 4.46 $\pm$ 2.60 & 7.03 $\pm$ 0.264 & 1.07 $\pm$ 0.123 \\\hline
ATLAS 13 TeV  & 5.77 $\pm$ 3.38 & 6.96 $\pm$ 0.265 & 1.12 $\pm$ 0.126  \\\hline
\end{tabular}}
\vspace{-20pt}
\end{table}


%
%
\bibliography{nsfthy,ismdproc}

%
%
%

\end{document}